\documentstyle[12pt,preprint]{aastex}  

\begin{document}

                           % The preamble begins here.
\title{A Survey for Satellites of Venus}  
\author{Scott S. Sheppard}    
\affil{Department of  Terrestrial Magnetism, Carnegie Institution of Washington, \\
5241 Broad Branch Rd. NW, Washington, DC 20015 \\ sheppard@dtm.ciw.edu}

\and

\author{Chadwick A. Trujillo}
\affil{Gemini Observatory \\
670 North A`ohoku Place, Hilo, HI 96720}

\begin{abstract}  % Produces abstract

We present a systematic survey for satellites of Venus using the
Baade-Magellan 6.5 meter telescope and IMACS wide-field CCD imager at
Las Campanas observatory in Chile.  In the outer portions of the Hill
sphere the search was sensitive to a limiting red magnitude of about
20.4, which corresponds to satellites with radii of a few hundred
meters when assuming an albedo of 0.1.  In the very inner portions of
the Hill sphere scattered light from Venus limited the detection to
satellites of about a kilometer or larger.  Although several main belt
asteroids were found, no satellites (moons) of Venus were detected.
  
\end{abstract}


\section{Introduction}

The Hill sphere radius, $r_H$, is the limiting radius for orbits of
planetary satellites in the presence of the Sun's gravitational field
and can be expressed as
\begin{equation}
r_H = a_p \left[\frac{m_p}{3M_{\odot}}\right]^{1/3}
\label{eq:hill}
\end{equation}
where $a_p$, $m_p$ and $M_{\odot}$ are the semi-major axis, mass of
the planet and mass of the Sun, respectively (Hill 1884; Innanen 1979;
Murray and Dermott 1999).  Hamilton and Krivov (1997) showed
analytically that the possible stability limit for satellites could be
closer to around $0.7 r_{H}$.  To date no known permanent satellite of
any planet has an orbit beyond $0.7 r_{H}$ from its primary.

Venus and Mercury are the only planets in our Solar System without any
known satellites.  Recent surveys of the giant planets have shown they
have extensive small outer satellite systems (Gladman et al. 1998,
2000 and 2001 ; Sheppard et al. 2003, 2005 and 2006; Holman et
al. 2004; Kavelaars et al. 2004).  These small outer irregular
satellites of the giant planets were likely captured from heliocentric
orbit near the end of the planet formation epoch (see Jewitt and
Haghighipour 2007 and Nicholson et al. 2008 for recent reviews on
irregular satellites).  Recent surveys show that the Terrestrial
planets Mars and Mercury do not have any outer satellites like the
giant planets (Sheppard et al. 2004; Nicholson and Gladman 2006;
Warell and Karlsson 2007).

The last published survey for satellites of Venus was performed using
photographic plates in 1956 (Kuiper 1961).  The 1956 satellite search
reached a limiting magnitude no better than about 16th in the R-band
for areas of the Hill sphere distant from the planet.  Thus the survey
could have detected objects larger than about 2.5 km in radius at
large distances from Venus.  Closer to the planet, the 1956 survey was
only able to obtain a limiting magnitude of about 14th, corresponding
to objects larger than about 6 km in radius.
  
The possible detection and discussion of a Venus satellite dates to at
least 1645 when F. Fontana mentioned the observation of a possible
Venus satellite.  Possible satellites of Venus were reported several
more times by many different and usually experienced observers
(including G. Cassini) in the late 1600's and 1700's (Blacklock 1868).
So many detections of a possible Venus satellite were made that
J. Lambert computed possible orbits and tables for the putative Venus
satellite in the late 18th century (Blacklock 1868; Anonymous 1884).
There has been no report of a Venus satellite since 1768 with many
notable astronomers such as W. Herschel and E. Barnard attempting
detection.  Hobbyists today have looked at Venus many times over with
telescopes that are more powerful than those from the 17th and 18th
century with no satellites reported.

Satellites have been invoked to explain Venus' retrograde rotation as
well as its impact crater record.  A Venus satellite (either
previously escaped or currently in-situ) could slow the rotation of
Venus through planet-satellite tidal friction, similar to the
Earth-Moon system (McCord 1968; Singer 1970; Kumar 1977; Donnison
1978; Malcuit and Winters 1995).  Bills (1992) notes that the pristine
state of most of Venus' impact craters is consistent with recent
tidal-induced decay of a swarm of small satellite fragments, possibly
from the destruction of a large parent satellite.  Alemi and Stevenson
(2006) suggest it is surprising that Venus has no satellites since its
very likely that Venus suffered several large impacts in the very
early Solar System.  These impacts would have a good chance of
creating a satellite, similar to how the Earth-Moon and Pluto-Charon
systems may have formed (Canup and Asphaug 2001; Canup 2005; Stern et
al. 2006).

Several authors have noted that Mercury and Venus may not have large
natural satellites as a consequence of strong solar gravitational
tides, which make large satellites unstable around the inner most
terrestrial planets (Counselman 1973; Ward and Reid 1973; Burns 1973;
Yokoyama 1999).  If Venus' slow rotation is primordial, Rawal (1986)
finds that Venus has trouble retaining all but the most distant and
smallest primordial satellites.  Satellites larger than a few km would
slowly spiral into the planet within the age of the Solar System.

Venus does have a few known quasi-satellites such as 2002 $\rm
VE_{68}$ (Mikkola et al. 2004).  Quasi-satellites are objects that
orbit the Sun in ellipses and have similar periods to the planet
(Wiegert et al. 2005).  In the planet's reference frame the object
resembles a retrograde elongated orbit around the planet.  These types
of orbits are usually destabilized over long periods of time by
gravitational interactions with neighboring planets (Mikkola et
al. 2006).  2002 $\rm VE_{68}$ is only expected to be a Venus
quasi-satellite for a few thousand years (Mikkola et al. 2004).  Any
primordial Venus Trojans are also unlikely to be stable for the age of
the Solar System (Tabachnik and Evans 2000; Brasser and Lehto 2002;
Scholl et al. 2005).

To date the Hill sphere of Venus has not been systematically surveyed
for possible small satellites with modern sensitive wide-field CCDs.
In order to constrain the presence of any small satellites of Venus a
deep CCD survey of the space around Venus was performed that is
several magnitudes more sensitive than any previously published
surveys for satellites of Venus.

\section{Observations}

Observations were made at the beginning of the night on UT October 7,
2005 with the Baade-Magellan 6.5 meter telescope at Las Campanas,
Chile.  Images were acquired in the R-band with the IMACS wide-field
CCD imager.  IMACS has eight $2048\times4096$ pixel CCDs with a pixel
scale of $0.20$ arcseconds per pixel.  The eight CCDs are arranged in
a box pattern with four above and four below and about 12 arcsecond
gaps between chips.  The field-of-view for IMACS is circular with a
radius of about 13.7 arcminutes giving an area of about 0.17 square
degrees.  This setup means the sky is vignetted at the extreme corners
of the outer CCDs and thus the corners are not used in the data
analysis.  Dithered twilight flat fields and biases were used to
reduce each image.  Landolt (1992) standards were used to
photometrically calibrate the data.  The night was clear and
photometric during all the observations.

During the observations Venus was near its highest point in the
Southern evening sky and thus between an airmass of 2.0 and 2.4.
Delivered image quality was between 1.2 and 1.3 arcseconds Full Width
at Half-Maximum (FWHM).  The apparent magnitude of Venus was about
-4.1 with a surface brightness of about 1.5 magnitudes per square
arcsecond.  Venus' angular diameter as seen on the sky was about 19
arcseconds with about 62 percent of Venus illuminated.  Venus'
geometric circumstances at the time of observations are shown in Table
1.

The most difficult aspect of a Venus satellite search is the large
amount of scattered light from Venus.  All images had a strong
gradient in the background because of this scattered light.  The light
gradient was removed by using the FMEDIAN task in IRAF.  The FMEDIAN
task replaces each pixel value with the median of the pixel values
around it.  For the Venus images each pixel had a box of $15\times15$
pixels used for the median.  The FMEDIAN image was then subtracted
from the original image.  Applying FMEDIAN to the images allows them
to be easily visually searched for moving objects.

Substituting the mass of Venus, $m_p = 4.87 \times 10^{24}$ kg, and
the mass of the Sun, $M_{\odot} = 1.99 \times 10^{30}$ kg, into
Eq.~\ref{eq:hill} yields a Venus Hill sphere radius of $r_H =
1.0\times 10^{6}$ km.  Using data from Table~1, we compute that Venus'
Hill radius on October 7, 2005 as seen from Earth was about 26.6
arcminutes (about 18.6 arcminutes for $0.7r_{H}$), very similar to the
full diameter of the field-of-view of one IMACS image.  In total the
Hill sphere of Venus covered about 0.62 square degrees in area as seen
from the Earth during the observations.

\begin{figure}
%\epsscale{0.4}
\centerline{\includegraphics[angle=0,width=\textwidth]{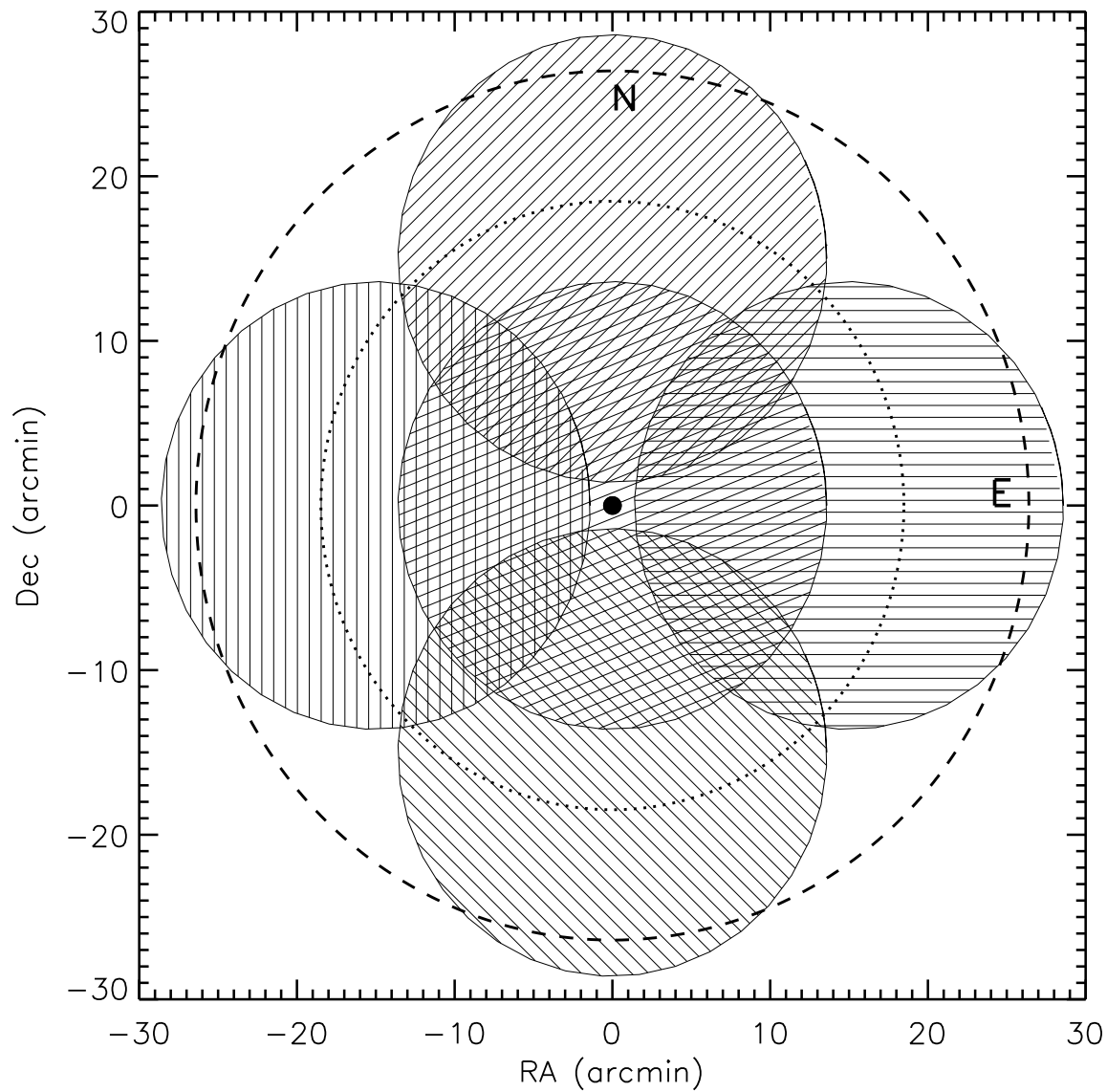}}
\caption{The area surveyed (shaded regions) around Venus (black
circle) for satellites using the Magellan-Baade 6.5 meter
telescope. Four fields (One North, South, East and West of Venus) were
imaged three times each around the planet on UT 2005 October 7.  An
additional field was imaged three times with Venus placed in the center
of the detector.  The dashed circle shows Venus' Hill sphere and the
dotted circle shows the theoretical outer limits for stable Venus
satellites ($0.7r_{H}$).}
\label{fig:venussurveyarea} 
\end{figure}

Figure~\ref{fig:venussurveyarea} illustrates the sky area surveyed
around Venus.  Data were collected at two epochs (Table 2).  First,
three 2 second exposures with Venus centered on the IMACS imager were
obtained.  These very short exposures prevented the CCDs from being
completely saturated and allowed satellites close to Venus to be
imaged.  The second part of the survey had Venus just offset to the
North, South, East and West of the IMACS field-of-view.  Three 10
second images were obtained of Venus in each of the offset positions.
On average there were about 4 minutes between exposures of Venus in
the same orientation.

\section{Analysis and Results}

The apparent Right Ascension (RA) and Declination (DEC) motion of
Venus during the observations is shown in Table 3.  Possible Venus
satellites would be expected to have similar apparent motions as Venus
(160 $\arcsec /hr$ in RA and -45 $\arcsec /hr$ in DEC).  Trailing
losses from the apparent motion of possible satellites was
insignificant since the few tenths of arcsecond trailing that would
occur in the 10 second images was much less than the 1.2 to 1.3
arcsecond image quality.  All known main belt asteroids in the survey
fields had apparent RA motion of less than 90 $\arcsec /hr$ in RA and
less than -20 $\arcsec /hr$ in DEC (Table 3).  No candidate satellites
of Venus (RA motion $> 90$ $\arcsec /hr$) were found through visually
blinking the survey fields.  Five main belt asteroids were detected in
the survey fields with apparent motions between about 50 and 80
arcseconds per hour in RA (see Table 3).

\begin{figure}
%\epsscale{0.4}
\centerline{\includegraphics[angle=90,width=\textwidth]{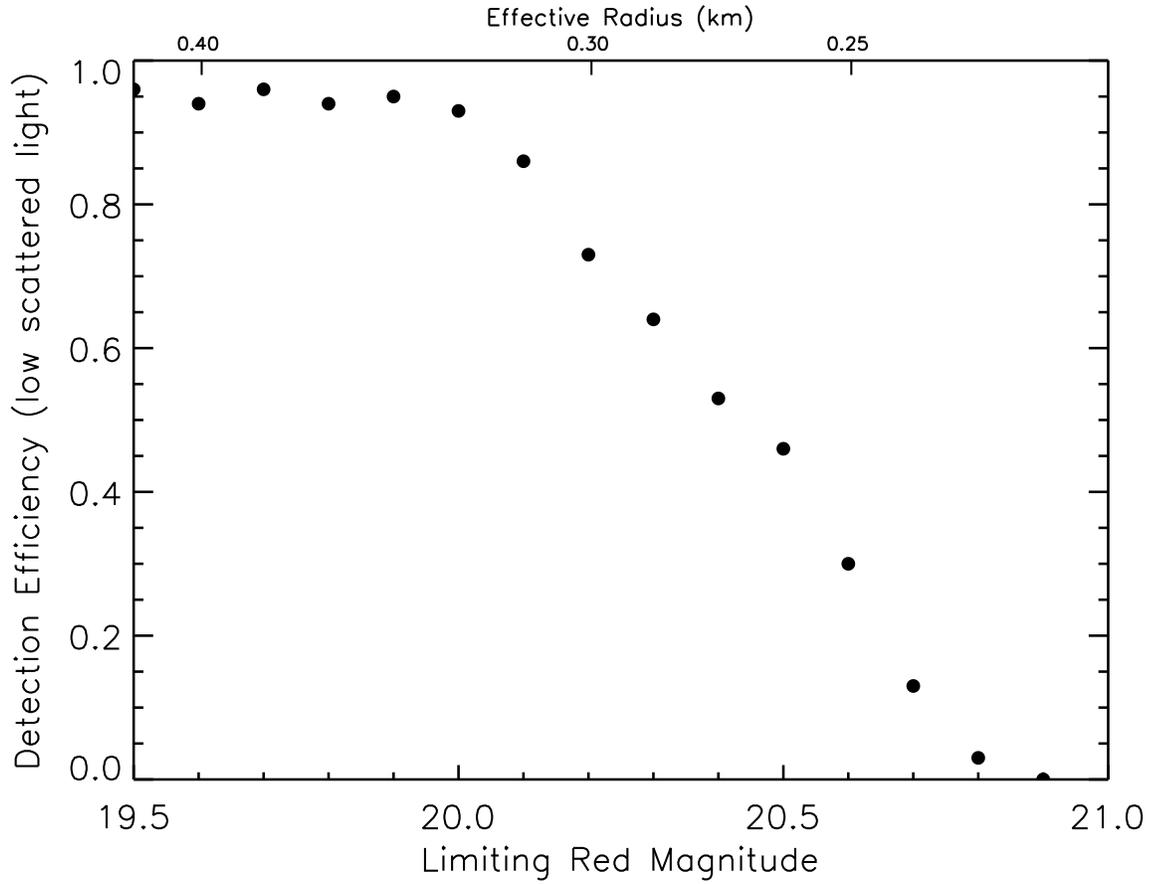}}
\caption{Detection efficiency of the artificially placed objects
during visual blinking of the fields.  The $50\%$ differential
detection efficiency is at about 20.4 mag.  This efficiency is valid
for the periphery of the survey area, where scattered light is
minimized.  The calculation of the effective radius assumes an albedo
of 0.1.}
\label{fig:detectioneff} 
\end{figure}

\begin{figure}
%\epsscale{0.4}
\centerline{\includegraphics[angle=90,width=\textwidth]{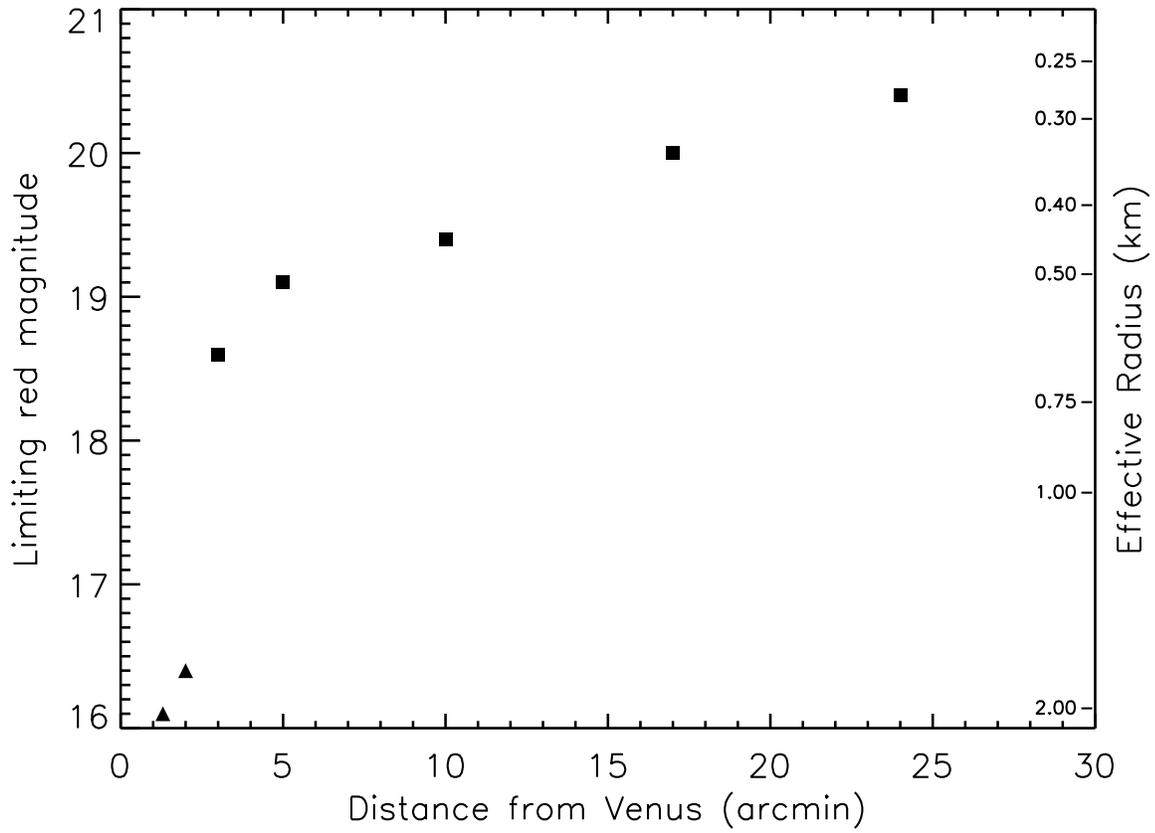}}
\caption{The $50\%$ detection efficiency of the survey versus distance
from Venus.  Squares represent the detection efficiency for the four
10 second fields with Venus offset from the center of the
detector. Triangles show the detection efficiency for the 2 second
field with Venus centered in the middle of the array.  The calculation
of the effective radius assumes an albedo of 0.1.}
\label{fig:effVenus} 
\end{figure}

\begin{figure}
%\epsscale{0.4}
\centerline{\includegraphics[angle=90,width=\textwidth]{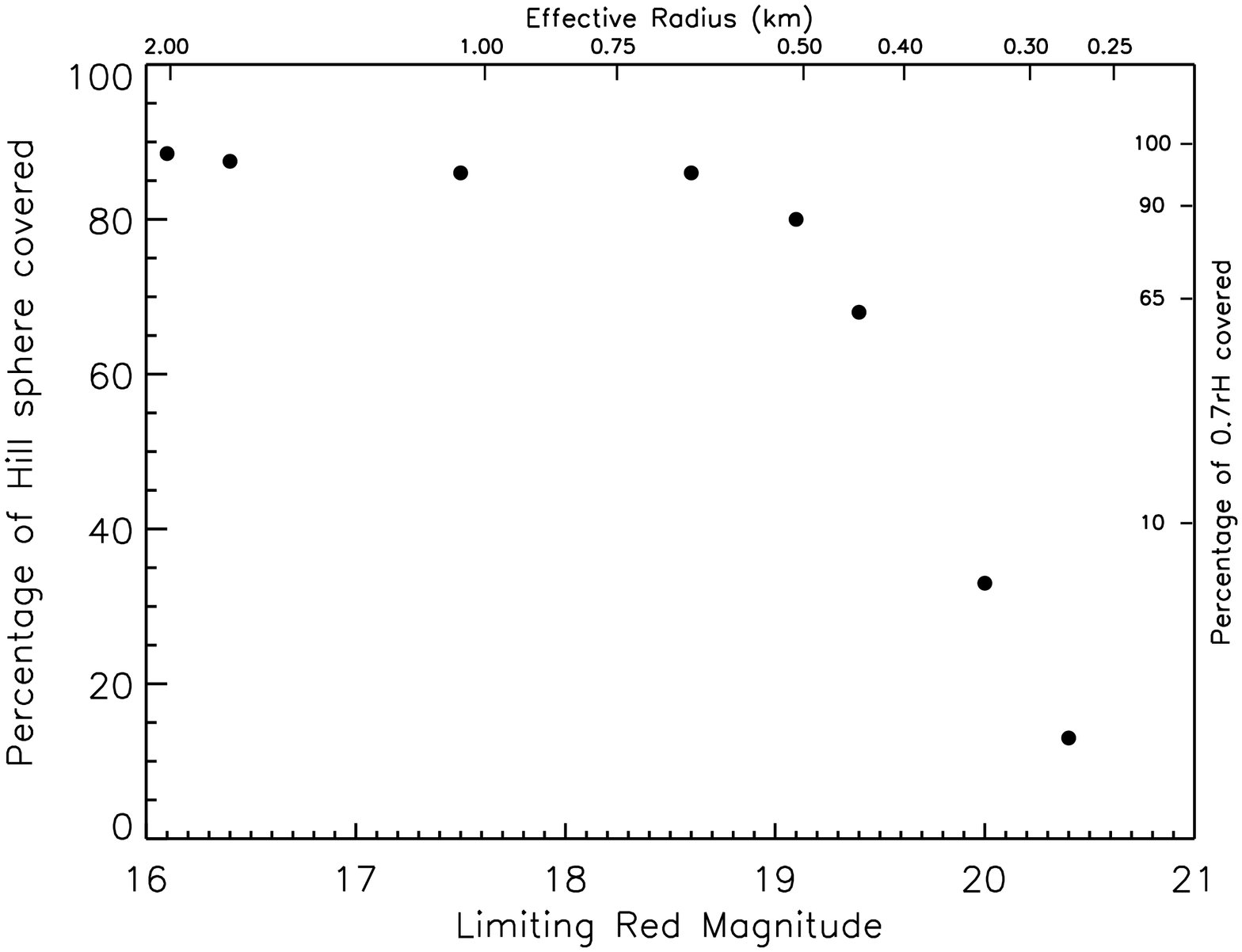}}
\caption{The completeness of the survey coverage of the Venus Hill
sphere versus the limiting red magnitude.  Because of the strong
scattered light near Venus a smaller percentage of the Hill sphere was
covered at fainter magnitudes.  The survey covered about $90\%$ of the
Venus Hill sphere and about $99\%$ of the Venus Hill sphere within
$0.7r_{H}$ or the theoretically stable region for satellites of Venus.
The percentage of the Hill sphere covered takes into account the
amount of area the survey covered at a particular limiting red
magnitude and the efficiency of detection at that magnitude.  The
calculation of the effective radius assumes an albedo of 0.1.}
\label{fig:comVenus} 
\end{figure}

The apparent red limiting magnitude, $m_{\mbox{\tiny{R}}}$, of the
survey was determined by placing artificially generated objects with
motions similar to that of Venus into the survey images.  The
artificial objects had R-magnitudes ranging between 14 and 21
magnitudes and were matched to the point spread function of the
images.  The $50\%$ differential detection efficiency was found to be
20.4 magnitudes for the artificial moving objects in the 10 second
images most distant from Venus (Fig.~\ref{fig:detectioneff}).
Scattered light was significant near Venus and the detection
efficiency versus distance from Venus is shown in
Fig.~\ref{fig:effVenus}.  The 10 second images became saturated around
3 arcminutes from Venus where the detection efficiency was around 18.6
magnitudes.  For the 2 second images with Venus centered on the array,
saturation occurred around 1.3 arminutes from Venus with a detection
efficiency at about 16.1 magnitudes (Fig.~\ref{fig:effVenus}).  About
$90\%$ of the Hill sphere around Venus was covered and about $99\%$ of
the Hill sphere within the theoretically stable area for satellites of
$0.7r_{H}$.  The percentage of the Venus Hill sphere covered per
limiting red magnitude is shown in Fig.~\ref{fig:comVenus}.

The corresponding radius limit, $r$, of an object to the apparent red
magnitude, $m_{\mbox{\tiny{R}}}$, can be found through
\begin{equation}
r = \left[ \frac{2.25\times 10^{16}R^{2}\Delta ^{2}}{p_{\mbox{\tiny{R}}}\phi
(\alpha)} \right]^{1/2} 10^{0.2(m_{\odot} - m_{\mbox{\tiny{R}}})}
\label{eq:appmag}
\end{equation}
in which $r$ is in km, $R$ is the heliocentric distance in AU,
$\Delta$ is the geocentric distance in AU, $m_{\odot}$ is the apparent
red magnitude of the sun ($-27.1$; Livingston (2000)), $p_{\mbox{\tiny{R}}}$ is the
geometric red albedo, and $\phi (\alpha)$ is the phase function in
which the phase angle $\alpha=0$ deg at opposition.  For an assumed
linear phase function the notation $\phi (\alpha) = 10^{-0.4 \beta
\alpha}$, where $\beta$ is the ``linear'' phase coefficient, is used.
Using data from Table 1 along with an S-type asteroid albedo of 0.1
and a linear phase coefficient of $\beta = 0.03$ mags per degree, as
found for Mercury and S-type asteroids (Veverka et al. 1988; Muinonen
et al. 2002), shows that 20.4 magnitudes corresponds to satellites
that are about 0.3 km (300 meters) in radius at Venus' observing
geometry.  Figures~\ref{fig:detectioneff} and~\ref{fig:effVenus} show
how the satellite radius corresponds to the survey's detection
efficiency.  This survey is a factor of about 50 deeper in flux than
the most recently published survey for satellites of Venus (Kuiper
1961).

\section{Summary}

No satellites of Venus down to about 0.3 km in radius were found in a
survey that covered about $90\%$ of the Hill sphere and $99\%$ of the
theoretically stable region for satellites of Venus.  The survey
improves the non detection of satellites around Venus by about a
factor of 50 over previously published work.  This result shows that
either Venus never acquired any satellites larger than about 1 km or
confirms that natural satellites larger than about 1 km around Venus
were unstable over the age of the Solar System.

\section*{Acknowledgments}

This paper includes data gathered with the 6.5 meter Magellan
Telescopes located at Las Campanas Observatory, Chile.  C.T. was
supported by the Gemini Observatory, which is operated by the
Association of Universities for Research in Astronomy, Inc., on behalf
of the international Gemini partnership of Argentina, Australia,
Brazil, Canada, Chile, the United Kingdom, and the United States of
America.

%{\it Facilities:} 
%\facility{Magellan:Baade (IMACS)}

\newpage

% This is a template LaTeX input file.  (Version of 17 August 1999)
% 
%
%\\ \colhead{} &
%\colhead{KBOs\tablenotemark{a}} & \colhead{}

%\documentstyle [aj_pt4]{article}    % Specifies the document style.

%\begin{document}

\begin{center}
\begin{deluxetable}{lccc}
%\small
\tablenum{1}
\tablewidth{6.0 in}
\tablecaption{Venus Geometrical Circumstances}
\tablecolumns{4}
\tablehead{
\colhead{UT Date} & \colhead{$R$}  & \colhead{$\Delta$} & \colhead{$\alpha$}  \\ \colhead{} &\colhead{(AU)} &\colhead{(AU)} &\colhead{(deg)} }  
\startdata
2005 October 7  & 0.7282 & 0.8782 & 76.3 \nl
\enddata
\tablenotetext{}{Quantities are the heliocentric distance ($R$), geocentric distance ($\Delta$) and phase angle ($\alpha$).}
\end{deluxetable}
\end{center}

%\end{document}             % End of document.

\newpage

% This is a template LaTeX input file.  (Version of 17 August 1999)
% 
%
%\\ \colhead{} &
%\colhead{KBOs\tablenotemark{a}} & \colhead{}

%\documentstyle [aj_pt4]{article}    % Specifies the document style.

%\begin{document}

\begin{center}
\begin{deluxetable}{lcccccc}
%\small
\tablenum{2}
\tablewidth{7.0 in}
\tablecaption{Venus Satellite Survey Fields}
\tablecolumns{7}
\tablehead{
\colhead{Field} & \colhead{RA(J2000)}  & \colhead{DEC(J2000)} & \colhead{EXP\tablenotemark{a}} & \colhead{Airmass} & \colhead{Filter} & \colhead{UT\tablenotemark{b}} \\ \colhead{} &\colhead{(hh:mm:ss)} &\colhead{(dd:mm:ss)} &\colhead{(sec)} & \colhead{}  & \colhead{}  & \colhead{(hh:mm:ss)} }  
\startdata
Center   &  15:43:48  &   -22:13:10  &   2   &   $2.0-2.1$   &   R    &     23:47:31/23:49:27/23:51:18  \nl
West     &  15:42:47  &   -22:13:17  &   10  &   $2.1-2.2$   &   R    &     23:53:12/23:59:43/00:03:00  \nl
East     &  15:44:56  &   -22:13:15  &   10  &   $2.1-2.2$   &   R    &     23:55:37/23:57:43/00:01:43  \nl
South    &  15:43:53  &   -22:58:24  &   10  &   $2.2-2.4$   &   R    &     00:05:38/00:09:25/00:18:58  \nl
North    &  15:43:54  &   -21:28:25  &   10  &   $2.2-2.4$   &   R    &     00:07:31/00:11:15/00:15:23  \nl
\enddata
\tablenotetext{a}{The exposure time of  each image.}
\tablenotetext{b}{The starting UT time of each image in the three image sequence.  Images with starting times of 23 hours were taken at the end of UT October 6, 2005 while images with starting times of 00 hours were taken at the beginning of UT October 7, 2005.}
\end{deluxetable}
\end{center}

%\end{document}             % End of document.

\newpage

% This is a template LaTeX input file.  (Version of 17 August 1999)
% 
%
%\\ \colhead{} &
%\colhead{KBOs\tablenotemark{a}} & \colhead{}

%\documentstyle [aj_pt4]{article}    % Specifies the document style.

%\begin{document}

\begin{center}
\begin{deluxetable}{lcccccccc}
%\small
\tablenum{3}
\tablewidth{7.0 in}
\tablecaption{Asteroids In The Venus Survey Fields}
\tablecolumns{9}
\tablehead{
\colhead{} & \multicolumn{2}{c}{Coordinates (J2000)} & \colhead{} & \multicolumn{2}{c}{Offsets} & \multicolumn{2}{c}{Motion} & \colhead{} \\ \colhead{Object} & \colhead{RA}  & \colhead{DEC} & \colhead{$m_{R}$} & \colhead{$\Delta$RA} & \colhead{$\Delta$DEC} & \colhead{dRA} & \colhead{dDEC} & \colhead{Detected} \\ \colhead{} &\colhead{(hh:mm:ss)} &\colhead{(dd:mm:ss)} &\colhead{(mag)} & \colhead{(arcmin)} & \colhead{(arcmin)} & \colhead{($\arcsec /hr$)} & \colhead{($\arcsec / hr$)} & \colhead{(exp/act)} }  
\startdata
Venus     & 15:43:52 & -22:13:30 & $-4.1$ & 0.0   & 0.0   &  $160$  & $-45$ &  Y/Y  \nl
44375     & 15:44:02 & -22:09:03 & $20.6$ & 2.2E  & 4.5N  &  $46 $  & $-10$ &  N/N  \nl
7408      & 15:43:10 & -22:14:03 & $18.3$ & 9.8W  & 0.5S  &  $80 $  & $-17$ &  Y/Y  \nl
76111     & 15:44:06 & -22:03:46 & $20.2$ & 3.3E  & 9.7N  &  $52 $  & $-13$ &  N/N  \nl
21748     & 15:43:55 & -22:24:01 & $19.8$ & 0.8E  & 10.5S &  $84 $  & $-17$ &  N/N  \nl
141541    & 15:44:35 & -22:22:14 & $20.5$ & 10.0E & 8.7S  &  $82 $  & $-20$ &  N/N  \nl
32536     & 15:44:07 & -22:26:42 & $18.0$ & 3.5E  & 13.2S &  $52 $  & $-7 $ &  Y/Y  \nl
2005 GS78 & 15:42:33 & -22:07:20 & $20.5$ & 18.2W & 6.2N  &  $54 $  & $-21$ &  N/N  \nl
51666     & 15:43:00 & -21:55:50 & $19.5$ & 12.0W & 17.7N &  $54 $  & $-17$ &  Y/Y  \nl
174368    & 15:44:26 & -21:53:28 & $20.5$ & 7.9E  & 20.0N &  $36 $  & $-15$ &  N/N  \nl
44717     & 15:42:21 & -22:22:19 & $19.0$ & 21.1W & 8.8S  &  $84 $  & $-19$ &  Y/Y  \nl
169311    & 15:42:13 & -22:16:21 & $20.5$ & 22.9W & 2.9S  &  $59 $  & $-11$ &  N/N  \nl
5710      & 15:42:16 & -22:20:52 & $19.0$ & 22.3W & 7.4S  &  $68 $  & $-14$ &  Y/N  \nl
109417    & 15:43:54 & -21:49:22 & $19.9$ & 0.4E  & 24.1N &  $49 $  & $-4 $ &  Y/Y  \nl
\enddata
\tablenotetext{}{Only asteroids in our survey fields and brighter than the maximum red magnitude survey limit, 20.8 magnitudes (see Fig. 2), on UT October 7, 2005 are presented. Because the survey's limiting magnitude becomes brighter closer to Venus the last column details if each object was expected (exp) to be detected and if it was actually (act) detected in our survey.  The asteroid information was obtained through the Minor Planet Center's MPChecker program.}
\end{deluxetable}
\end{center}

%\end{document}             % End of document.

\newpage

%\newpage

%\newpage

%\newpage

\begin{references}

\reference{Ano84} Anonymous, 1884. The problematical satellite of Venus. The Observatory 7, 222-226.

\reference{Bil92} Bills, B., 1992. Venus: Satellite orbital decay, ephemeral ring formation, and subsequent crater production. Geophysical Research Letters 19, 1025-1028.

\reference{Bla68} Blacklock, A., 1868. The Satellite of Venus. Astronomical register 6, 196-197.

\reference{Bra02} Brasser, R, Lehto, H., 2002. The role of secular resonances on trojans of the terrestrial planets. MNRAS 334, 241-247.

\reference{Bur73} Burns, J., 1973. Solar system- Why are there no satellites for the inner planets? Nature Phys. Sci. 242, 23.

\reference{Can01} Canup, R. and Asphaug, E., 2001. Origin of the Moon in a giant impact near the end of the Earth's formation. Nature 412, 708-712.

\reference{Can05} Canup, R., 2005. A giant impact origin of Pluto-Charon. Science 307, 546-550.

\reference{Cou73} Counselman, C., 1973. Outcomes of tidal evolution. ApJ 180, 307-316.

\reference{Don78} Donnison, J., 1978. The escape of natural satellites from Mercury and Venus. Ap\&SS 59, 499-501.

\reference{gla98} Gladman, B., Nicholson, P., Burns, J., Kavelaars, J., Marsden, B., Williams, G., Offutt, W., 1998. Discovery of two distant irregular moons of Uranus. Nature 392, 897-899.

\reference{gla00} Gladman, B., Kavelaars, J., Holman, M., Petit, J., Scholl, H., Nicholson, P., Burns, J., 2000. Note: The discovery of Uranus XIX, XX, and XXI. Icarus 147, 320-324.

\reference{gla01} Gladman, B., Kavelaars, J., Holman, M., Nicholson, P., et al. 2001. Discovery of 12 satellites of Saturn exhibiting orbital clustering. Nature 412, 163-166.

\reference{Ham97} Hamilton, D. and Krivov, A., 1997. Dynamics of distant moons of asteroids. Icarus 128, 241-249.

\reference{Hil78} Hill, G., 1884. Mr. G. W. Hill's paper on lunar theory. MNRAS 44, 194-196. 

\reference{Hol03} Holman, M., Kavelaars, J., Grav, T., Gladman, B. et al. 2004, Discovery of five irregular moons of Neptune. Nature 430, 865-867.

\reference{Inn79} Innanen, K., 1979. The limiting radii of direct and retrograde satellite orbits, with applications to the solar system and the stellar systems. AJ 84, 960-963.

\reference{Hag07} Jewitt, D., Haghighipour, N., 2007. Irregular satellites of the planets: Products of capture in the early solar system. ARA\&A 45, 261-295.

\reference{Kav04} Kavelaars, J., Holman, M., Grav, T., Milisavljevic, D., et al. 2004. The discovery of faint irregular satellites of Uranus. Icarus 169, 474-481.

\reference{Kui61} Kuiper, G., 1961. Limits of completeness. in: Planets and Satellites, eds. G. Kuiper and B. Middlehurst, (University of Chicago Press; Chicago) pp. 575-591.

\reference{Kum77} Kumar, S., 1977. The escape of natural satellites from Mercury and Venus. Ap\&SS 51, 235-238. 

\reference{Lan92} Landolt, A., 1992. UBVRI photometric standard stars in the magnitude range 11.5-16.0 around the celestial equator. AJ 104, 340-371.

\reference{Liv00} Livingston, W., 2000. Sun. in: Allen's astrophysical quantities, eds. A. Cox, (AIP Press; New York) pp. 339-380.

\reference{Mal95} Malcuit, R., Winters, R., 1995. Numerical simulation of retrograde gravitational capture of a satellite by Venus: Implications for the thermal history of the planet. LPI 26, 885-886.

\reference{Mcc68} McCord, T., 1968. The loss of retrograde satellites in the solar system. J. Geophys. Res. 73, 1497-1500.

\reference{Mik04} Mikkola, S., Brasser, R., Wiegert, P., Innanen, K., 2004. Asteroid 2002 $\rm VE_{68}$, a quasi-satellite of Venus. MNRAS 351, L63-L65.

\reference{Mik06} Mikkola, S., Innanen, K., Wiegert, P., Connors, M., Brasser, R. 2006. Stability limits for the quasi-satellite orbit. MNRAS 369, 15-24.

\reference{Mui02} Muinonen, K., Piironen, J., Shkuratov, Y., Ovcharenko, A., Clark, B., 2002. Asteroid photometric and polarimetric phase effects. in: Asteroids III, eds. W. Bottke Jr., Cellino, A., Paolicchi, P. and R. Binzel, (The University of Arizona Press; Tucson) pp. 123-138.

\reference{Mur99} Murray, C., Dermott, S., 1999. Solar System dynamics. (Cambridge University Press; Cambridge)

\reference{Nic06} Nicholson, P., Gladman, B., 2006. Satellite searches at Pluto and Mars. Icarus 181, 218-222.

\reference{Nic08} Nicholson, P., Cuk, M., Sheppard, S., Nesvorny, D., Johnson, T., 2008. Irregular satellites of the giant planets. in: The Solar System Beyond Neptune, eds. M. Barucci, H. Boehnhardt, D. Cruikshank and A. Morbidelli, (The University of Arizona Press; Tucson) pp. 411-424.

\reference{Raw86} Rawal, J., 1986. Possible satellites of Mercury and Venus. EM\&P 36, 135-138.

\reference{Rus16} Russell, H., 1916. On the albedo of the planets and their satellites. ApJ 43, 173-195.

\reference{Sch05} Scholl, H., Marzari, F., Tricarico, P., 2005. The instability of Venus Trojans. AJ 130, 2912-2915.

\reference{She03} Sheppard, S., Jewitt, D., 2003. An abundant population of small irregular satellites around Jupiter. Nature 423, 261-263.

\reference{She04} Sheppard, S., Jewitt, D., Kleyna, J., 2004. A survey for outer satellites of Mars: Limits to completeness. AJ 128, 2542-2546.

\reference{She05} Sheppard, S., Jewitt, D., Kleyna, J., 2005. An ultradeep survey for irregular satellites of Uranus: Limits to completeness. AJ 129, 518-525.

\reference{She06} Sheppard, S., Jewitt, D., Kleyna, J., 2006. A survey for ``normal'' irregular satellites around Neptune: Limits to completeness. AJ 132, 171-176. 

\reference{Sin70} Singer, S., 1970. How did Venus lose its angular momentum? Science 170, 1196-1198.

\reference{Ste06} Stern, S.A., Weaver, H., Steffl, A. et al. 2006. A giant impact origin for Pluto's small moons and satellite multiplicity in the Kuiper belt. Nature, 439, 946-948.

\reference{Tab00} Tabachnik, S., Evans, N., 2000. Asteroids in the inner solar system - I. Existence. MNRAS 319, 63-79.

\reference{Vev88} Veverka, J., Helfenstein, P., Hapke, B., Goguen, J., 1988. Photometry and polarimetry of Mercury. in: Mercury, eds. F. Vilas, C. Chapman and M. Matthews, (The University of Arizona Press; Tucson) pp. 37-58.

\reference{War73} Ward, W., Reid, M., 1973. Solar tidal friction and satellite loss. MNRAS 164, 21-32.

\reference{War07} Warell, J., Karlsson, O., 2007. A search for natural satellites of Mercury. P\&SS 55, 2037-2041.

\reference{Wie00} Wiegert, P., Connors, M., Brasser, R., Mikkola, S., Stacy, G., Innanen, K., 2005. Sleeping with an elephant: Asteroids that share a planet's orbit. J. Royal Astro. Soc. Can. 99, 145.

\reference{Yok99} Yokoyama, T., 1999. Dynamics of some fictitious satellites of Venus and Mars. Planetary and Space Science 47, 619-627.


\end{references}
\end{document}